# Magnetism based on nitrate-nitrate interactions: The cases of $LiNO_3$, $K_{0.5}Rb_{0.5}NO_3$, $Ca(NO_3)_2$ and $C(NH_2)_3NO_3$

Na Du,[‡1] Xintian Wang,[‡1] Ruo Tong Wang,[2] Enting Xu,[1] Yu Ying Zhu,[2] Yan Zhao,[1] Peng Ren,*[1] Fei Yen*[1]

[1]School of Science, Harbin Institute of Technology, Shenzhen, University Town, Shenzhen, Guangdong 518055, P. R. China.
[2]School of Mechanical Engineering and Automation, Harbin Institute of Technology, Shenzhen, University Town, Shenzhen, Guangdong 518055, P. R. China.
Correspondence: fyen at hit.edu.cn; renpeng at hit.edu.cn.

**Abstract:** Long-range magnetic ordering of the orbital motion of oxygen atoms within $NO_3^-$ cations is identified from experimental measurements of the magnetic susceptibility $\chi(T)$ in $LiNO_3$, $Ca(NO_3)_2$, $K_{0.5}Rb_{0.5}NO_3$ and $C(NH_2)_3NO_3$ at their respective order-disorder, solid-solid phase transitions $T_N$. The observed sharp changes in $\chi(T)$ and accompanying hysteretic behavior indicate the phase transitions to be first order. A model employing the law of conservation of angular momentum is used to explain why the librations between neighboring $NO_3^-$ become geared below $T_N$. Since the periodic motions involve concerted motion of net charges, the associated magnetic moments of the $NO_3^-$ ions indirectly establish an antiferromagnetic structure below $T_N$. Our findings identify a previously unidentified type of molecular interaction which may be exploited to further increase the enthalpy of the widely-popular hydrated salts employed as energy storage devices.

# 1. Introduction

When a substance undergoes a first-order phase transition, energy is absorbed/released in the latent heat of transition. If the latent heat can be stored and later retrieved and consumed, then the substance may be designated as a "phase change material" (PCM). The practicality of PCMs chiefly depends on the amount of heat absorbed/released in the intended operating temperature range and its transition temperature among many other factors [1-3]. Microscopically, the amount of latent heat across a phase transition is largely dependent on the change in entropy of the material according to how much the individual ions become ordered or disordered. At solid-liquid and solid-solid phase transitions the ordering is usually translational and orientational, respectively. Often in solid-solid phase transitions of salts containing polyatomic ions, the change in entropy is anomalously higher than the expected amount according to all the accountable orientational configurations [4-6]. Recently, we identified that such excess changes in entropy may be due to a previously unaccounted contribution stemming from temporal ordering of the ionic reorientations which also happens to exhibit magnetic behavior [7].

In the case of nitrate salts, the changes in entropy at some of the phase transitions are anomalously larger than their counterpart halide salts [8,9]. The change in entropy $\Delta S$ spreads over across the phase transition so the enthalpy values near room temperature are rather large making the series of nitrate salts popular candidates for energy storage applications [1-3,10-12]. Simply accounting for the ordering of the spatial orientations of the nitrate cations $NO_3^-$ does not seem to be sufficient because $\Delta S$ is larger than $R \ln 2$ (where $R$ is the natural gas constant) at the solid-solid phase transitions [6,13] so we hypothesized that temporal ordering is also involved. Given the planar nature and symmetry of the nitrate cation, they can easily reorient within their planes with little disturbance to the lattice [14,15]. Even at liquid helium temperatures, the in-plane libration frequencies of many nitrate salts vary from 15 to 32 cm$^{-1}$ [16]. Hence, we picture that at high temperatures, each cation librates independently with respect to its adjacent cations, *i.e.* it randomly rotates back and forth within its plane by discrete angles of ±$q$ confined by a wide potential barrier. However, below a threshold temperature $T_N$, usually the order-disorder phase transition point, the librations become geared with respect to each other so if an $NO_3^-$ rotates by +$q$, then its adjacent neighbors do so by –$q$. Accounting of this temporal ordering mechanism can explain the excess changes in entropy across the solid-solid phase transitions. Incidentally, as an $NO_3^-$ rotates by $q$, an extremely weak magnetic field is generated which we can approximately equate it to a dipolar magnetic moment **μ**. If the network of $NO_3^-$ below $T_N$ indeed becomes geared, then an antiferromagnetic structure of **μ** should emerge (Fig. 1). Macroscopically, antiferromagnetic ordering of negatively charged species causes the magnetic susceptibility to exhibit a pronounced change at the transition temperature. In the cases when all electrons are paired so the system is non-magnetic (diamagnetic), a detection of a sharp change in the magnetic susceptibility at the order-disorder phase transition of a nitrate salt would provide direct evidence of an onset of $NO_3^-$–$NO_3^-$ magnetic interactions.

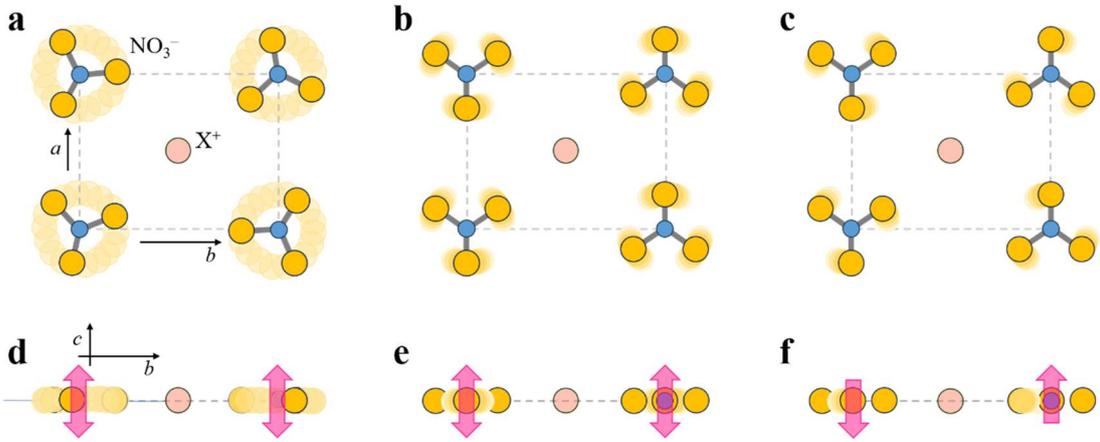

**Fig. 1.** Schematic diagram depicting the motions and degree of ordering of planar $NO_3^-$ ions along the *ab*-planes (upper panels) when (a) disordered; (b) orientationally ordered; and (c) temporal ordered. The bottom panels (d)–(f) show associated magnetic moments **μ** generated by orbital motion of $O^{2/3-}$ ions. The current consensus is that the system experiences the process from (a) to (b) and (d) to (e) during a disordering-ordering process. In contrast, we suggest the system undergoes the process from (a) to (c) and (d) to (f).

In the following, we report on the experimental measurements of the magnetic susceptibilities as functions of temperature of the diamagnetic salts $LiNO_3$, $Ca(NO_3)_2$, $K_{0.5}Rb_{0.5}NO_3$ and $C(NH_2)_3NO_3$; they represent nitrate salts with cations that are spherical univalent/divalent, mixed and polyatomic. We then provide a simple model explaining why temporal ordering of the $NO_3^-$ librations occurs at the phase transitions and how this process, as an indirect consequence, induces antiferromagnetic ordering at the molecular scale.

## 2. Materials and methods

Anhydrous lithium nitrate $LiNO_3$ was purchased from Energy Chemicals, Inc. The samples (tiny crystals crushed into powder and pressed into disc pellets) were first annealed at 390 K for 2 hours to remove any remaining $H_2O$ prior to each measurement. Calcium nitrate $Ca(NO_3)_2$ was obtained by reacting nitric acid with calcium carbonate. The samples were re-crystallized three times, crushed into powder and pressed into pellets for the measurements. The samples were also annealed at 390 K before the measurements to remove the tetrahydrate phase. Mixed potassium rubidium nitrate $K_{0.5}Rb_{0.5}NO_3$ samples were obtained by mixing an equimolar ratio of $KNO_3$ and $RbNO_3$ in deionized $H_2O$ to form a solution and allowed to slowly evaporate at room temperature. The ratio was confirmed by the additivity rule of the molar magnetic susceptibility method [17]. Guanidinium nitrate crystals were grown from the slow evaporation of a solution comprised of its reagent and deionized $H_2O$. To show the high purity of the samples and the absence of paramagnetic impurities, the expected and experimentally obtained diamagnetic susceptibilities are compared in the Supplementary Material file. The magnetic susceptibility measurements were carried out with the VSM option of a PPMS (Physical Property Measurement System) manufactured by Quantum Design, Inc. The raw data of all the figures along with a more detailed explanation of the experimental procedure is available in the Supplementary Material file.

## 3. Results and discussion

### 3.1 Lithium nitrate $LiNO_3$

Lithium nitrate $LiNO_3$ is the simplest nitrate salt; its structure remains trigonal from room temperature up to its melting point [18]. According to electrical resistance measurements a transition occurs at 263 K [19]. Figure 2 shows the molar magnetic susceptibility with respect to temperature $\chi(T)$ of polycrystalline LiNO$_3$ under an applied magnetic field of $H = 1$ T. The experimentally obtained value of $\chi(300\ \text{K})$ was $-19.5\times10^{-6}$ cm$^3$/mol, which is in good agreement with the expected value of $-19.9\times10^{-6}$ cm$^3$/mol according to addition of Pascal's constants [20]. At 264.7 K, $\chi(T)$ increased in magnitude to $-24.7\times10^{-6}$ cm$^3$/mol (by ~27%). Upon warming, the phase transition occurred at 302.7 K and $\chi(T)$ reverted back to nearly its original value. Such a drastic change in $\chi(T)$ and a presence of hysteresis can only signify a first-order phase transition involving the ordering of magnetic moments. In the case of negatively charged species, if the change in slope in $\chi(T)$ is positive then the ordering is antiferromagnetic. Since the electrons are paired in LiNO$_3$ (as it is diamagnetic), the pronounced changes in the measured $\chi(T)$ curves of LiNO$_3$ is solid evidence of antiferromagnetic ordering of the NO$_3^-$ species.

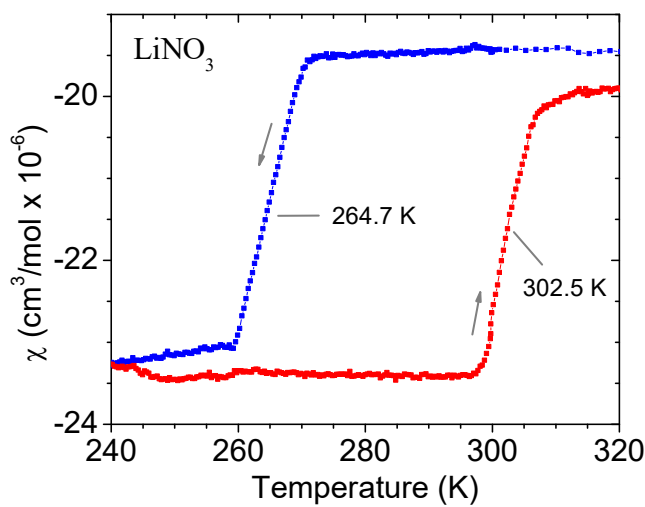

**Fig. 2.** Molar magnetic susceptibility as a function of temperature $\chi(T)$ of polycrystalline lithium nitrate LiNO$_3$ under applied magnetic field of $H = 1$ T. Arrows indicate the cooling and warming curves under sweeping rate of 0.5 K/min. Data in all of the figures are available in the Supplementary Material file.

### 3.2 Calcium nitrate Ca(NO$_3$)$_2$

Calcium nitrate Ca(NO$_3$)$_2$ exhibits a wide λ-peak anomaly in its heat capacity in the range between 280 K and 292 K (with its peak at 287.8 K) [21,22]. Experiments from dielectric constant measurements concurred on the presence of such a solid-solid phase transition just below room temperature [23]. The crystal structure of the room temperature phase of $Ca(NO_3)_2$ is cubic [22] but that of the low temperature phase remains unknown. Given that the $NO_3^-$ of most divalent nitrate salts also exhibit in-plane librations at low temperatures [16], it was natural for us to investigate whether such order-disorder phase transition pertains to an onset-offset gearing of the $NO_3^-$. Figure 3 shows $\chi(T)$ of polycrystalline $Ca(NO_3)_2$. The most remarkable feature is the large hysteretic region in between 245 and 318 K suggesting the phase transition to be first order. However, the region from 289 K to 241 K during cooling (and 293 K to 318 K during warming), where the slope of $\chi(T)$ is positive and continuously changes by over 14% in value, also indicates the phase transition to occur gradually similar to $NaNO_3$ [24]. A possible mechanism underlying the current case is the occurrence of an intermediate phase where the gearing (or interlocking) of the $NO_3^-$ librations require a finite amount of time to complete. We note that most alkaline nitrate salts exhibit order-disorder phase transitions, but their transition temperatures are much higher than 400 K, the upper limit of our measurement system, so we encourage researchers to investigate their magnetic properties as this should bring forth more information regarding the lattice dynamics of the system.

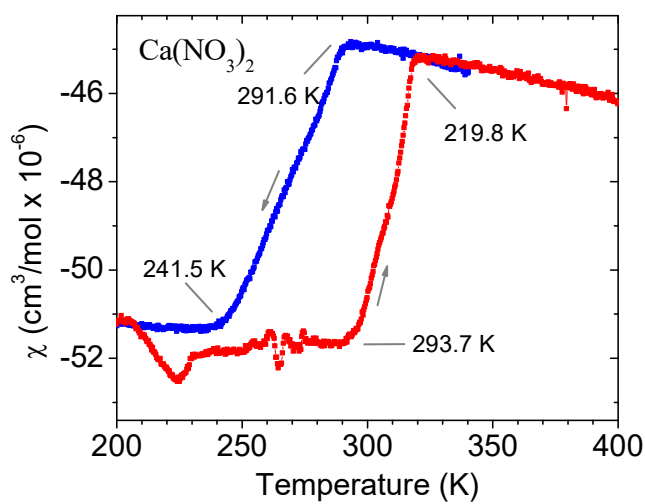

**Fig. 3.** $\chi(T)$ of calcium nitrate $Ca(NO_3)_2$ in polycrystalline form; sweeping rate was 1 K/min.

### 3.3 Potassium nitrate $KNO_3$

Potassium nitrate $KNO_3$ exhibits a phase transition near 402 K when warming from its room temperature ordered phase II to the high temperature disordered phase I.

During cooling, a metastable ferroelectric phase III exists in the region 397–387 K before the system transforms back to phase II [25,26]. With a 50% substitution of $K^+$ by $Rb^+$, the phase transition temperatures II–I, I–III and III–II decrease to around 387, 365 and 355 K, respectively [27,28]. We investigated $K_{0.5}Rb_{0.5}NO_3$ because all of the transition points occur below 400 K. Figure 4 shows $\chi(T)$ of single crystalline $K_{0.5}Rb_{0.5}NO_3$ measured along the 001 direction which is also that of the spontaneous polarization. A rather sharp change of 8.3% occurred in the span of 384.5±1.6 K upon heating indicating the II-I phase transition taking place. During cooling, a similar step-like feature was also observed, but the I-II phase transition seemed to have been interrupted by a presence of a metastable state starting at 376.8 K which can only be attributed to the ferroelectric III phase. Only near 370.0 K did the system seemed to have reverted back to phase II. If there were no changes in the magnetic interactions between $NO_3^-$ ions, then $\chi(T)$ would remain independent of temperature; this is true even upon crossing any structural phase transition involving no magnetic ordering. Hence, it is evident that there is also an intimate connection between the ferroelectricity of phase III and the dynamics of $NO_3^-$ ions. Most other univalent nitrate salts such as $Na^+$, $Rb^+$, $Cs^+$, $Tl^+$ and $Ag^+$ exhibit solid-solid, order-disorder phase transitions at high temperatures so it is pertinent to investigate their $\chi(T)$ up to their respective melting points in the attempts of better understand their lattice dynamics.

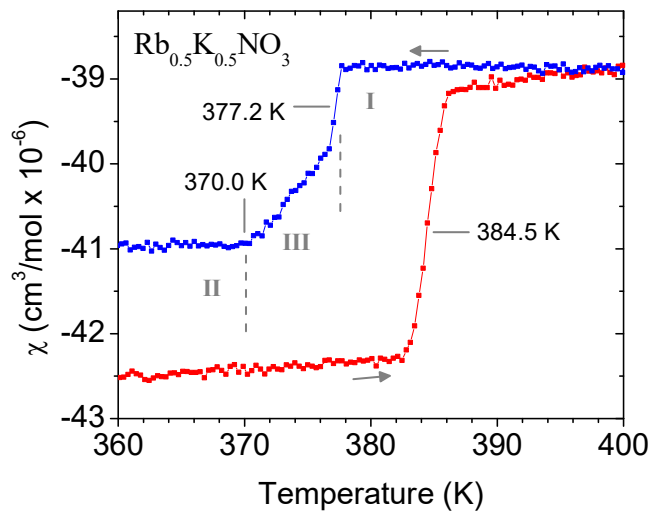

**Fig. 4.** $\chi(T)$ of potassium rubidium nitrate $K_{0.5}Rb_{0.5}NO_3$ in single crystalline form. $\chi(T)$ and $H$ were along the 001 direction of the crystal and the sweeping rate was 1 K/min.

### 3.4 Guanidinium nitrate C(NH₂)₃NO₃

The order-disorder phase transition of guanidinium nitrate C(NH$_2$)$_3$NO$_3$ has been well-studied as it involves an extremely large change of its lattice constant despite the low temperature and high temperature phases remaining monoclinic [29,30]. Existing literature from various types of measurements pinpoint the phase transition to occur at 276 K and 296 K during cooling and warming, respectively [31]. Figure 5 displays χ(*T*) of guanidinium nitrate along the 010 orientation of a single crystalline sample. During the first cooling run, no pronounced anomaly was observed. Upon warming, a sharp increase by 9.4% in χ(*T*) occurred at 296.4 K. The second cooling run showed a sharp decrease in χ(*T*) at 276.3 K. All subsequent runs exhibited pronounced changes near the same transition points but systematically decreasing in magnitude. This can be explained by the fact that the crystals were grown at 295 K so it seems that the NO$_3^-$ were in an ordered state at the start of the experiments. Then, with each cycle the crystal cracked more each time the phase transition was crossed which decreased the number of NO$_3^-$–NO$_3^-$ interactions.

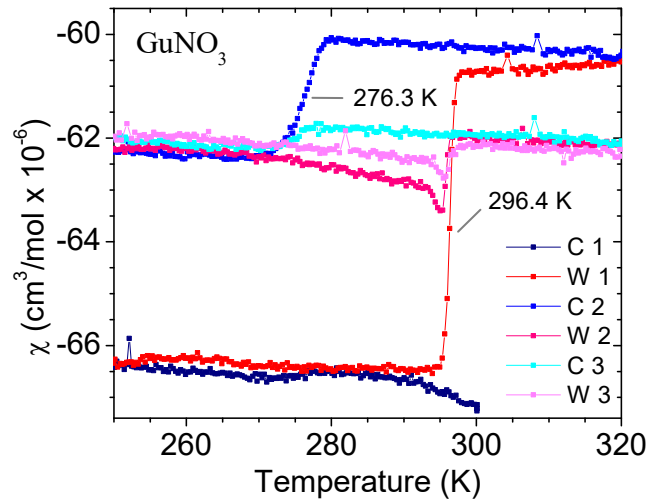

**Fig. 5.** χ(*T*) along the 010 orientation of a guanidinium nitrate crystal measured under 1 K/min.

### 3.5 Discussion

To explain the reason why an order-disorder phase transition takes place, we make use of the law of conservation of angular momentum. Mathematically, this is expressed as the total angular momentum $L_{system}$ of a closed system remaining constant with respect to time:

$$\frac{d}{dt}(L_{system}) = 0$$

In crystalline solids with polyatomic ions, apart from phonons [32,33] $\nu_n(T)$ contributing to $L_{system}$ (where $n = 1, 2, …$), the orbital motion of the non-spherical ions also contributes an $mv_nr$ component, where $m$ and $v_n$ are, respectively, the mass and velocity of the non-stationary atoms and $r$ the average radius of the orbits. For the case of $NO_3^-$, $m$ would be the masses of the three oxygen atoms, $r$ the average distance between the O and N atoms and $v_n$ the speed of the O atoms when the entire anion rotates by an angle $q$. Since $r$, $v_n$ and $\nu_n$ are all dependent on temperature $T$, Eq. 1 can be expanded to become:

$$\frac{d}{dt}(m\boldsymbol{v}_1\boldsymbol{r}(T) + m\boldsymbol{v}_2\boldsymbol{r}(T) + \cdots + \nu_1(T) + \nu_2(T) + \cdots) = 0$$

We can start by only considering two $NO_3^-$ (mediated by a rigid cation) in the ground state where $mv_nr \gg \nu_n$ (minimal phonon contribution). Hence, from Eq. 2, $m\boldsymbol{v}_1\boldsymbol{r} \approx -m\boldsymbol{v}_2\boldsymbol{r}$, meaning that if one $NO_3^-$ rotates by $q$, then the other $NO_3^-$ must rotate by $-q$, i.e. the two cations are geared, or "ordered". As $T$ is increased, the phonon contribution becomes significantly enough that adjacent $NO_3^-$ no longer need to remain geared because there are other channels to conserve angular momentum. Since the oxygen atoms (with negative net charges) at the extremities trace out current loops, each rotating $NO_3^-$ carries with itself an equivalent magnetic moment $\mu$ [34-36]. Hence, $\mu$ is arranged antiferromagnetically in the ordered phase while behaving paramagnetically in the disordered phase. The experimentally measured macroscopic magnetization is sensitive to changes of all of the individual $\mu$ but not $\nu_n$. This explains the sharp changes observed in $\chi(T)$ at the order-disorder phase transitions for the present cases of $LiNO_3$, $Ca(NO_3)_2$, $K_{0.5}Rb_{0.5}NO_3$ and $C(NH_2)_3NO_3$.

## 4. Conclusions

To conclude, we investigated the magnetic properties of several nitrate salts and found evidence of antiferromagnetic ordering of the magnetic moments arising from librating $NO_3^-$. We suggest the formation of the antiferromagnetic structures to be indirect manifestations of the $NO_3^-$ rotors becoming geared due to coarse graining of conservation of angular momentum. Our findings provide an explanation to the excess change in entropy encountered in most nitrate salts as originating from previously not accounting the gearing mechanism of the $NO_3^-$. Consequently, it is worthy to investigate whether *partial* isotope exchange of the $^{16}O$ atoms by $^{18}O$ would increase the magnetic entropy of a system which in turn may *further* increase their heat capacities, thus the enthalpies and energy storage capacities of the currently widely-employed hydrated salts as phase change materials. Lastly, the $NO_3^-$ in divalent nitrates and $CO_3^{2-}$ in carbonates also librate at liquid helium temperatures [16], so we also expect to observe magnetic ordering based on oxygen atom orbitals at the solid-solid phase transitions present in many organic & inorganic nitrates, carbonates and other salts possessing oxyanions.

**CRediT authorship contributions statement**

**Na Du** prepared the $LiNO_3$ samples, grew the $K_{0.5}Rb_{0.5}NO_3$ crystals, performed most of the magnetic susceptibility measurements, and drew Figures 2-6. **Xin Tian Wang** performed XRD analysis on all of the samples. **Ruo Tong Wang** assisted with the magnetic susceptibility measurements, data analysis and plotting of the graphs. **En Ting Xu** grew the guanidinium nitrate crystals and helped synthesize calcium nitrate. **Yu Ying Zhu** and **Yan Zhao** assisted with the magnetic susceptibility measurements, crystal growth process and drawings. **Peng Ren** partly supervised the preparation and quality of the samples. **Fei Yen** designed the project, supervised the experiments and drafted the manuscript.

**Data Availability**

All data are included in the Supplementary Information file.

**Declaration of Competing Interest:**

The authors declare no competing financial interests.